# Synthesis and physical properties of $Ca_{1-x}RE_xFeAs_2$ with $RE$ = La ~ Gd

Alberto Sala[1,2]*, Hiroyuki Yakita[1], Hiraku Ogino[1], Tomoyuki Okada[1], Akiyasu Yamamoto[1], Kohji Kishio[1], Shigeyuki Ishida[3], Akira Iyo[3], Hiroshi Eisaki[3], Masaya Fujioka[4], YoshihikoTakano[4], Marina Putti[2] and Jun-ichi Shimoyama[1]

[1]*Department of Applied Chemistry, The University of Tokyo, 7-3-1 Hongo, Bunkyo, Tokyo 113 8656, Japan*
[2]*University of Genova and CNR-SPIN, Via Dodecaneso, 16146, Genova, Italy*
[3]*National Institute of Advanced Industrial Science and Technology (AIST), Tsukuba, Ibaraki 305-8565, Japan*
[4]*National Institute for Materials Science* (NIMS)*, 1-2-1 Sengen, Tsukuba, Ibaraki 305-0047, Japan*

E-mail: alberto.sala@spin.cnr.it

Synthesis of a series of layered iron arsenides $Ca_{1-x}RE_xFeAs_2$ (112) was attempted by heating at 1000°C under a high-pressure of 2 GPa. The 112 phase successfully forms with $RE$ = La, Ce, Nd, Sm, Eu and Gd, while Tb, Dy and Ho substituted and $RE$ free samples does not contain the 112 phase. The Ce, Nd, Sm, Eu and Gd doped $Ca_{1-x}RE_xFeAs_2$ are new compounds. All of them exhibit superconducting transition except for the Ce doped sample. The behaviour of the critical temperature, with the RE ionic radii have been investigated.

Since the discovery of the iron-based superconductors with a transition temperature ($T_c$) of 26 K in LaFeAs(O,F)[1], there has been a relevant focus towards synthesizing novel iron pnictide as; 1111 type - $RE$FeAs(O,F) ($RE$ = rare earth elements), 122 type - $AE$Fe$_2$As$_2$ ($AE$ = alkaline earth metals), 111 type - $AE$FeAs, 11 type – Fe$Ch$ ($Ch$ = chalcogenides elements)[2-4]. Later other structures with more complex blocking layers have been discovered, such as compounds having perovskite-type oxide layer (ex. Fe$_2$P$_2$Sr$_4$Sc$_2$O$_6$) and platinum based Ca$_{10}$(Pt$_3$As$_8$)(Fe$_2$As$_2$) and Ca$_{10}$(Pt$_4$As$_8$)(Fe$_2$As$_2$)$_5$[5-8]. In order to gain more light on the mechanism of superconductivity in iron-based compounds and hopefully to achieve further enhancement of $T_c$ the discovery of new iron-based superconductor families is crucial.

Recently 112 type iron-based superconductors Ca$_{1-x}RE_x$FeAs$_2$ has been reported for $RE$ = La and Pr[9-10]. The space group of the compounds is a monoclinic $P2_1/m$, and the structure is composed of two Ca($RE$) planes, anti-fluorite Fe$_2$As$_2$ layers, and As$_2$ zigzag chain layers.

The highest $T_c$ reported for this new family is 43 K for Sb-doped Ca$_{0.85}$La$_{0.15}$FeAs$_2$ sample, estimated by magnetic measurements[11]. This compound has several interesting features such as As$^-$ in As$_2$ zigzag chain layers, and further studies are needed to investigate more detailed superconducting properties. In addition, incorporation of La and Pr in CaFeAs$_2$ suggested that other $RE$ elements can partially substitute for the Ca-site. Since the ionic size of $RE$ affects the local crystal structure at the Fe$_2$As$_2$ layers, change of superconducting properties is expected by changing $RE$ elements. In this paper, we report the synthesis and characterization of Ca$_{1-x}RE_x$FeAs$_2$ compounds, with $RE$ free and $RE$ = La ~ Ho except for Pm.

All samples were synthesized by high pressure synthesis (HPS) using wedge-type cubic-anvil high-pressure apparatus (TRY engineering Co. Ltd.) starting from FeAs(3N), $RE$As(3N), Ca(2N), and As(4N) powders. Preparation of the samples was carried out in an argon-filled glove box. Powder mixtures were prepared following the nominal composition of CaFeAs$_2$ or Ca$_{0.85}RE_{0.15}$FeAs$_2$ which shows the best superconducting properties[11], subsequently the powder were pelletized and reacted in boron nitride crucible at 1000°C for 1 h under 2 GPa. Phases identification was carried out by XRD measurements using RIGAKU Ultima-IV diffractometer and intensity data were collected in the $2\theta$ range of 5°-80° at a step of 0.02° using Cu-$K_\alpha$ radiation. Observation of microstructure and compositional analysis were performed on plate like single crystals using a SEM (Hitachi High-Technologies TM3000) equipped with EDX (Oxford Instruments SwiftED 3000).

Magnetic susceptibility was measured by a SQUID magnetometer (Quantum Design MPMS-XL5s). Electrical resistivity was measured by the AC four-point-probe method using Physical Property Measurement System (Quantum Design).

Powder XRD analyses have been performed in all the synthetized samples. They revealed that 112 phase was successfully formed in $RE$ doped samples with $RE$ = La ~ Gd by HPS at 1000°C for 1 h under 2 GPa, while FeAs and $FeAs_2$ were always detected as impurity phases. In addition, very small amount of $CaFe_2As_2$ impurity was contained in the Eu doped sample. On the other hand, any diffraction peaks due to the 112 phase were not found in the XRD patterns of Tb, Dy and Ho doped samples and a $RE$ free sample.
In our preliminary study to synthesize $RE$ doped $CaFeAs_2$ samples at ambient pressure, the 112 phase was found to form only by substitutions of Ce, Pr and Nd. By applying HPS we succeeded to introduce Sm, Eu and Gd in the 112 phase showing that high pressure is crucial for including smaller $RE$ ions in this phase. This suggests that by increasing the synthesis pressure above 2 GPa, it could make possible the incorporation of smaller $RE$ ions, such as Tb, Dy, Ho and Y, in the 112 lattice.

Small silvery crystals of 112 phase were obtained from opaque black bulk samples with $RE$ = La~Eu. Fig. 1 shows surface XRD patterns for the plate like single crystals of La and Nd doped 112. Crystals of approximately 200 × 200 × 10 µm$^3$, as shown in the inset of Fig. 1, were chosen for the XRD analysis. As clearly seen, the pattern of Nd doped 112 compound show (00$l$) peaks shifting towards higher 2$\theta$ values in respect to La doped 112 compound: this indicates a decrease of the lattice parameters, which reflect the smaller $RE$ ion ionic radii of Nd in respect to La.

The EDX analysis on different crystals extracted from the bulk samples has been performed. The atomic percentage of Fe and As is close to 0.25 and 0.5, respectively, as expected from the nominal composition of (Ca,$RE$)$FeAs_2$. Contrarily, the $RE$ concentration has been found to differ from the nominal composition. An average composition of $RE$ per Ca site in La, Ce, Nd, Sm and Eu doped samples are 25, 26, 23, 11, and 17%, respectively. This analysis has not been performed on Gd doped compound since this sample was composed of small crystals and it could not possible to extract plate-like crystals with large surface area to obtain a reliable EDX signal. The $RE$ concentration in the crystals is always higher than nominal composition except for the Sm doped crystals. Kudo et al.[11] reported that $T_c$ of La doped $CaFeAs_2$ monotonically decreases with an increase in La concentration over 15% at the Ca site and, hence, our samples except for Sm and Eu doped ones that are close to the optimal doping, can be considered to be slightly over-doped.

Magnetic susceptibility curves measured from 2 K to 50 K for a series of 112 samples are shown in Fig. 2. That of Pr doped sample synthesizes at ambient pressure reported by Yakita *et al.*[10] was also added for comparison. Note that bulk samples containing several small crystals as well as plate-like crystals were examined in these measurements, because each plate-like crystal is too small to perform accurate measurements. Superconducting transitions are clearly observed in the La Pr, Nd, Sm, Eu and Gd doped samples, while the Ce doped sample does not show any trace of superconducting transition down to 2 K, instead a paramagnetic like behaviour is observed. The transitions are quite broadened probably due to the not homogeneous distribution of the dopant in the selected crystals. In the case of Eu and Gd, which present sharper transitions a rough evaluation of the shielding volume fraction at 2 K was around 17 % and 15 % respectively: The estimated shielding volume fraction for the La doped sample is very low, close to 0.7%. This value is extremely low in comparison to the other $(Ca,La)FeAs_2$ reported in literatures[9-11]. We suppose that the superconductivity in our sample is depressed by the to the La overdoping (25%).

The $T_{c\text{-mag}}$ is determined by the cross point of the linear interpolation of the normal and superconducting state of magnetic susceptibility. The evaluated $T_{c\text{-mag}}$ of the La, Pr, Nd, Sm, Eu and Gd doped samples are 24.5, 13.2, 11.9, 11.6, 9.3, and 12.6 K, respectively. $T_{c\text{-mag}}$ progressively decreases with increasing the atomic number of the substituted *RE*, the Gd doped sample showed relatively high $T_{c\text{-mag}}$, probably because of lower doping level.

The results of resistivity measurements performed on bulk samples are shown in Fig. 3. All the samples exhibit a metallic behaviour. Consistent with the magnetic measurements, the Ce doped sample did not show superconducting transition. In contrast, the La, Pr, Nd, Sm, Eu and Gd doped samples showed clear superconducting transitions, though transitions are very broad probably due to inhomogeneous *RE* concentration of the superconducting phases in the bulk samples and poor grain connectivity originated in coexisting impurity phases. The Eu doped sample, similarly to the magnetic measurements, presents the sharper transition. The resistivity onset transition $T_{c\text{-res}}$, evaluated using the intersection point method, for the La, Pr, Nd, Sm, Eu and Gd doped samples are = 22.7, 24.6, 17.9, 25.5, 13.2 and 22.8 K, respectively. These values are higher than those obtained by magnetization measurements and this is a clear indication of the inhomogeneity of the samples. As it is well known, the resistivity measurement is sensitive to the local high $T_c$ regions, whereas magnetization to the average critical temperature of the sample. In this case no clear relationship between $T_{c\text{-res}}$ and *RE* atomic number can be found. An interesting feature is flex points observed for all *RE* doped samples between 50 K to 80 K. The features can be regarded as an intrinsic behaviour of the 112 phase associated to magnetic ordering or structural transition, though more detailed structural and magnetic investigation are necessary to clarify this point.

Distances between Fe planes $d_{Fe-Fe}$ evaluated from the XRD patterns of single crystals and the $T_{c-mag}$ are summarized in Fig. 4 as a function of the ionic radii of the *RE* ions. The $d_{Fe-Fe}$ values correspond to $c\sin\beta$, where $c$ is the $c$-axis length and $\beta$ is the angle in between the $a$- and $c$-axes in the monoclinic system. With decreasing the ionic size of the substituted *RE* ions $d_{Fe-Fe}$ values decrease.

The exception is the Eu doped sample, which showed a large $d_{Fe-Fe}$ value out of the main trend. This can be the evidence of the presence of $Eu^{2+}$ ions, because of larger ionic radii of $Eu^{2+}$(1.25 Å) in comparison to $Eu^{3+}$ (1.066 Å). On the other hand, the $d_{Fe-Fe}$ value of the Ce doped sample is in trend with the other *RE* doped samples, indicating that the presence of smaller $Ce^{4+}$ ions seems to be excluded. For the Gd doped sample, $d_{Fe-Fe}$ value was impossible to calculate because plate like crystals were not grown and its powder XRD pattern included many peaks due to large amount of impurities, such as FeAs and $FeAs_2$, which overlap peaks of 112 phase.

In order to correlate the $T_c$ with the ionic radii of the *RE* doping atoms the $T_{c-mag}$ is reported in Figure 4. It's possible to observe a decreasing of $T_{c-mag}$ with decreasing the ionic radii of the doping atoms, even if the variation is substantial only passing from the La sample to the others *RE*. Magnetic susceptibility superconducting transitions are broad and this suggests: high inhomogeneity of the samples and a strong dependence of $T_c$ from the doping level. At the present condition the doping level control is not precise and improvement of synthesis procedure are required in order to refine this analysis.

We successfully synthesized a series of *RE* doped $CaFeAs_2$ compounds with *RE* = La~Gd by applying HPS method. All the samples showed clear superconducting behaviour, except for the Ce doped sample. Relatively low $T_c$'s observed in our samples and actual *RE* concentration largely differed from nominal ones which suggests the precise control of *RE* concentration in the samples is crucial for optimization of superconducting properties as well as for understanding the determining factors of $T_c$ in this new superconductor system. The absence of superconductivity in the Ce doped samples is not related to a size effect of the *RE* ions or presence of $Ce^{4+}$, since $d_{Fe-Fe}$ value of Ce doped sample is in between the La and Pr doped samples. In addition, the resistivity curves of all samples showed flex points that deserve more detailed analysis. Further additional studies using higher quality single crystals are required to clarify origins of those unconventional behaviours.

**Acknowledgments**

This work is supported by: Japan Society for the Promotion of Science (JSPS), and FP7 European Union-Japan project SUPER-IRON (grant agreement No. 283204).

**References**

1) Y. Kamihara, T. Watanabe, M. Hirano and H. Hosono, J. Am. Chem. Soc. **130**, 3296 (2008).

2) M. Rotter, M. Tegel, D. Johrendt, Phys. Rev. Lett. **101**, 107006 (2008).

3) S. Matsuishi, Y. Inoue, T. Nomura, T. Yanagi, M. Hirano and H. Hosono, J. Am. Chem. Soc. **130**, 14428 (2008).

4) F. C. Hsu, J. Y. Luo, K. W. Yeh, T. K. Chen, T. W. Huangh, P. M. Wu, Y. C. Lee, Y. L. Huang, Y. Y. Chu, D. C. Yan and M. K. Wu, Proc. Natl. Acad. Sci. U.S.A. **23**, 14262 (2008).

5) H. Ogino, Y. Matsumura, Y. Katsura, K. Ushiyama, S. Horii, K. Kishio and J. Shimoyama, Supercond. Sci. Technol. **22**, 075008 (2009).

6) S. Kakiya, K. Kudo, Y. Nishikubo, K. Oku, E. Nishibori, H. Sawa, T. Yamamoto, T. Nozaka, M. Nohara, J. Phys. Soc. Jpn. **80**, 093704 (2011).

7) N. Ni, J. M. Allred, B. C. Chan and R. J. Cava, Proc. Natl. Acad. Sci. U.S.A. **108**, E1019 (2011).

8) C. Löhnert, T. Stürzer, M. Tegel, R. Frankovsky, G. Friederichs and D. Johrendt, Angew. Chem., Int. Ed. **50**, 9195 (2011).

9) N. Katayama, K. Kudo, S. Onari, T. Mizukami, K. Sugawara, Y. Sugiyama, Y. Kitahama, K. Iba, K. Fujimura, N. Nishimoto, M. Nohara, and H. Sawa, J. Phys. Soc. Jpn. **82**, 123702 (2013).

10) H. Yakita, H. Ogino, T. Okada, A. Yamamoto, K. Kishio, T. Tohei, Y. Ikuhara, Y. Gotoh, H. Fujihisa, Kunimitsu Kataoka, H. Eisaki, and J. Shimoyama, J. Am. Chem. Soc., **136**, 846 (2014).

11) K. Kudo, T. Mizukami, Y. Kitahama, D. Mitsuoka, K. Iba, K. Fujimura, N. Nishimoto, Y. Hiraoka and M. Nohara, J. Phys. Soc. Jpn. **83**, 025001 (2014)

**Figure Captions**

**Fig. 1.** Surface XRD patterns and optical image of a plate-like crystal, shown in the figure's inset, of (Ca,*RE*)FeAs$_2$ with *RE* = La and Nd.

**Fig. 2** ZFC magnetic susceptibility measured at H = 10 Oe for Ca$_{0.85}$*RE*$_{0.15}$FeAs$_2$ bulk samples, in the inset magnification of the La doped sample.

**Fig. 3.** Temperature dependence of the normalized resistivity for Ca$_{0.85}$*RE*$_{0.15}$FeAs$_2$ bulk samples, the inset shows the magnification of the superconducting transition.

**Fig. 4.** $d_{\text{Fe-Fe}}$ value and $T_{\text{c - mag}}$ as a function of the ionic radii of the *RE*$^{3+}$ ions in a coordination number (C.N.) of 8, for the (Ca,*RE*)FeAs$_2$ samples. Dotted and dashed line lines are guide for eyes only.

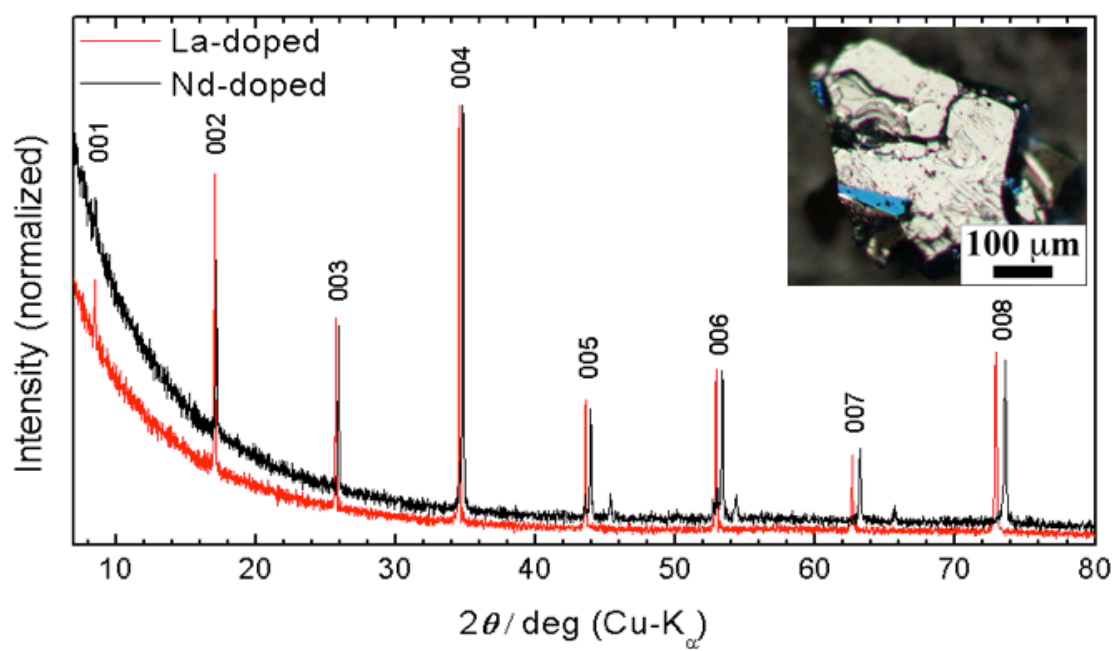

Fig. 1.

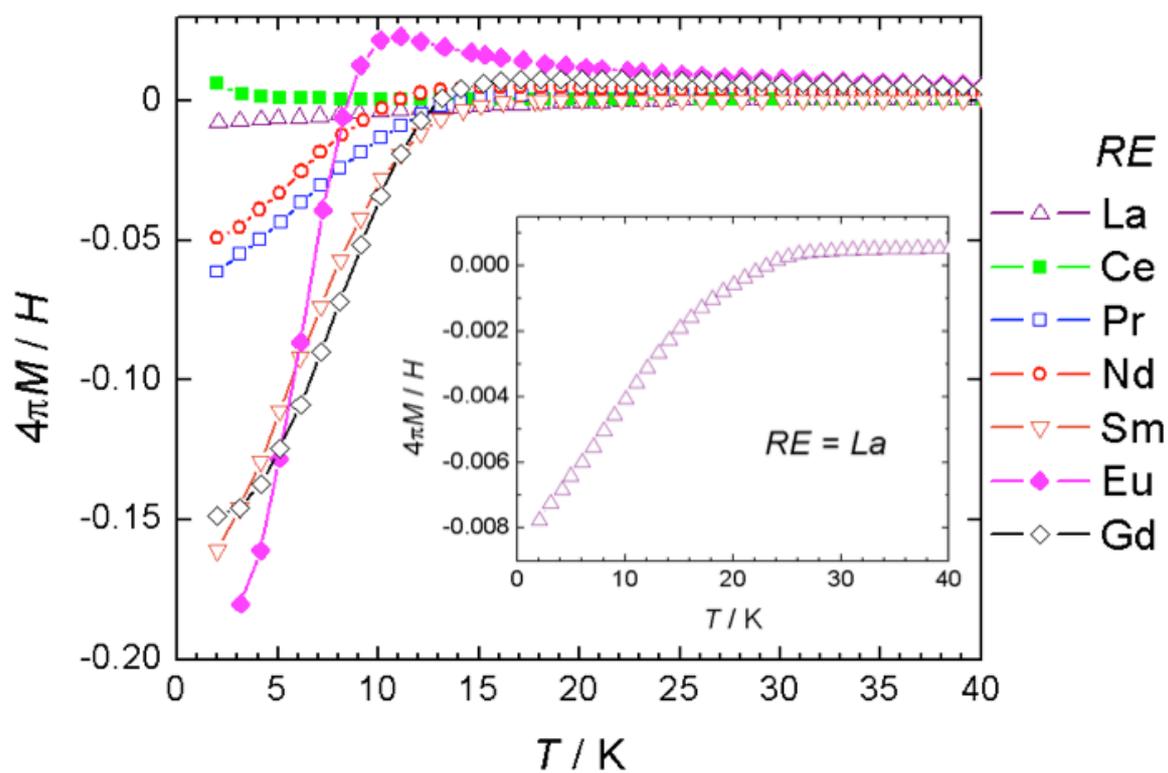

Fig. 2.

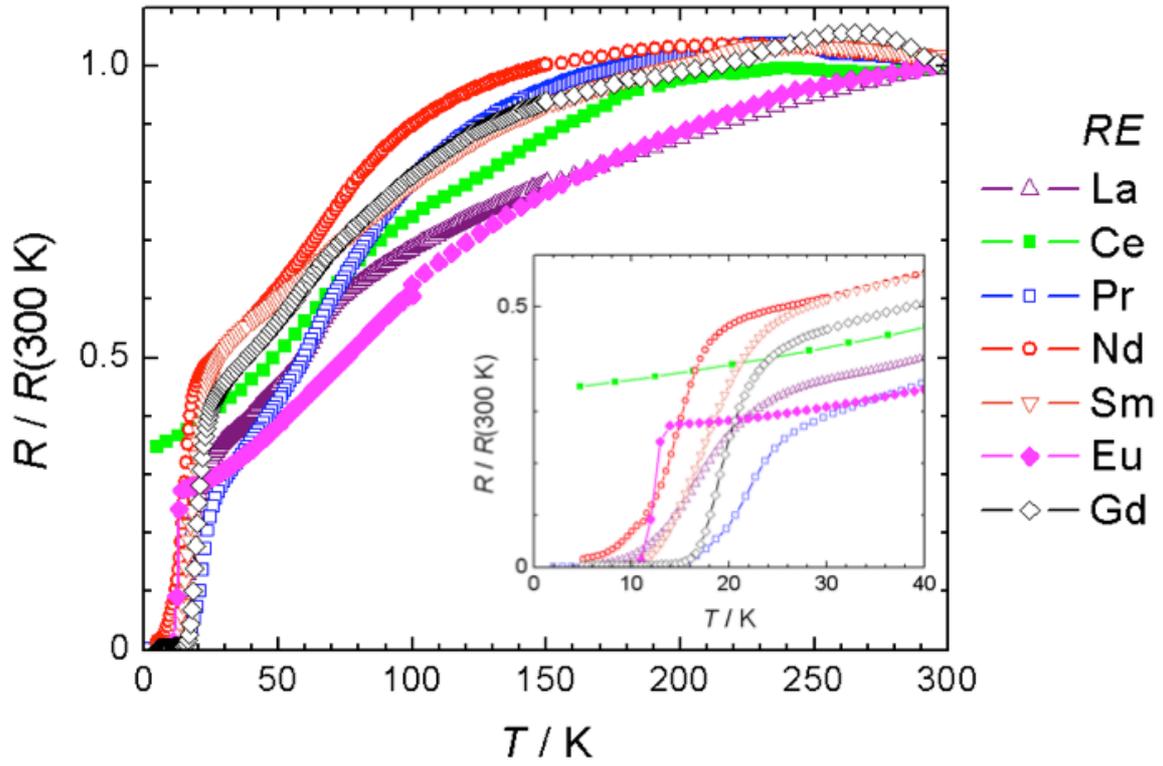

Fig. 3.

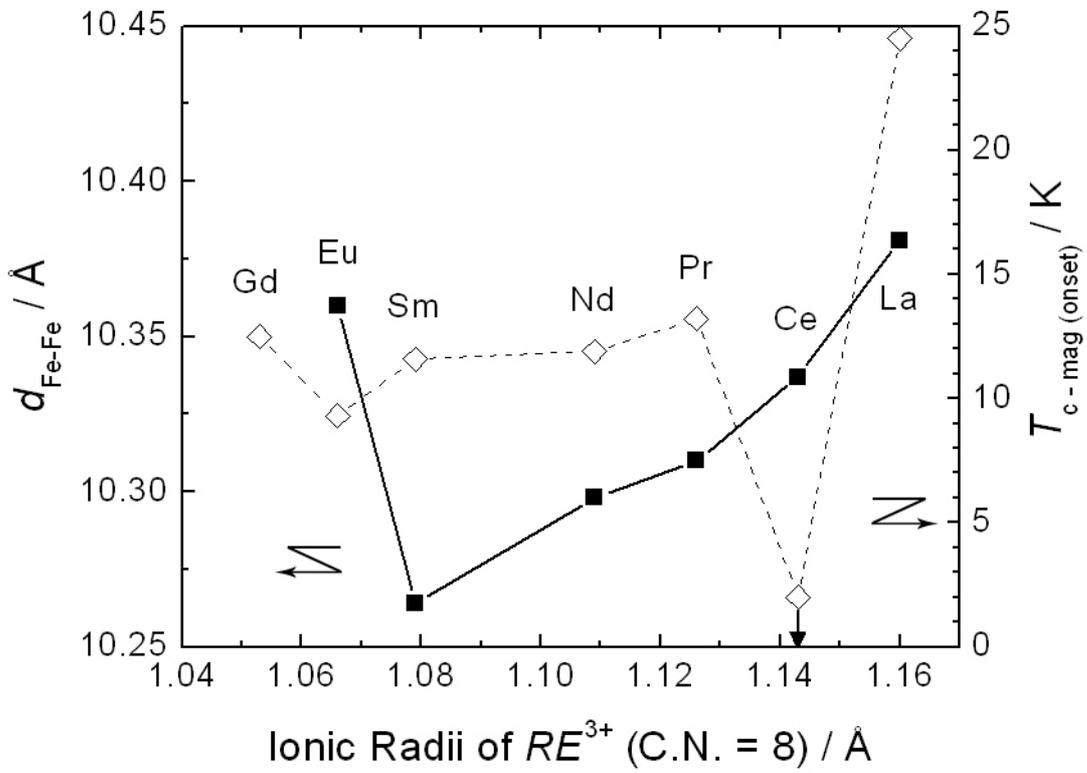

Fig. 4.